\documentclass[preprint,prb]{revtex4}%
\usepackage{amssymb}
\usepackage{amsmath}
\usepackage{graphicx}
\usepackage{amsfonts}%
\providecommand{\U}[1]{\protect\rule{.1in}{.1in}}
\begin{document}
\title[ ]{Faraday Induction and the Current Carriers in a Circuit}
\author{Timothy H. Boyer}
\affiliation{Department of Physics, City College of the City University of New York, New
York, New York 10031}

\begin{abstract}
In this article, it is pointed out that Faraday induction can be treated from
an untraditional, particle-based point of view. \ The electromagnetic fields
of Faraday induction can be calculated explicitly from approximate
point-charge fields derived from the Li\'{e}nard-Wiechert expressions or from
the Darwin Lagrangian. Thus the electric fields of electrostatics, the
magnetic fields of magnetostatics, and the electric fields of Faraday
induction can all be regarded as arising from charged particles. \ Some
aspects of electromagnetic induction are explored for a hypothetical circuit
consisting of point charges which move frictionlessly in a circular orbit.
\ For a small number of particles in the circuit (or for non-interacting
particles), the induced electromagnetic fields depend upon the mass and charge
of the current carriers while energy is transferred to the kinetic energy of
the particles. \ However, for an interacting multiparticle circuit, the mutual
electromagnetic interactions between the particles dominate the behavior so
that the induced electric field cancels the inducing force per unit charge,
the mass and charge of the individual current carriers become irrelevant, and
energy goes into magnetic energy. \ 

\end{abstract}
\maketitle

\section{Introduction}

When students are asked what causes the electric field in a parallel-plate
capacitor, the response involves charges on the capacitor plates. \ Also,
students say that the magnetic field in a solenoid is due to the currents in
the solenoid winding. \ But when asked for the cause of the Faraday induction
fields in a solenoid with changing currents, the student response is that the
induction field is due to a changing magnetic field, not that it is due to the
acceleration fields of the charges in the solenoid winding. The student view
reflects what is emphasized in the standard electromagnetism
textbooks.\cite{Griffiths},\cite{Jackson} \ Indeed, although C. G.
Darwin\cite{D-metals} computed induction fields from accelerating charges in
the 1930s, today it is rare to have a physicist report that the Faraday
induction field arises from the acceleration of charges.\cite{Essen}%
,\cite{Hill}

In this article, we wish to broaden the perspective on Faraday induction by
reviewing some aspects of the particle point of view. \ We treat the induction
fields as arising from the electromagnetic fields of point charges as derived
from the Li\'{e}nard-Wiechert expressions or from the Darwin
Lagrangian.\ First we mention the Li\'{e}nard-Wiechert\cite{J664} form taken
by the electric and magnetic fields of a point charge in general motion.
\ Then we turn to the low-velocity-small-distance approximation derived in the
1940 textbook by Page and Adams.\cite{Electrodynamics} \ This approximate form
for the electromagnetic fields is the same as that obtained from the Darwin
Lagrangian\cite{Darwin} of 1920. \ Here in the present article, the
approximated fields are used to treat Faraday induction in detail for a
hypothetical circuit consisting of point charged particles moving
frictionlessly on a circular ring. \ This circuit provides a rough
approximation to that of a thin wire of finite thickness which is bent into a
circular loop. \ For small numbers of particles (or for noninteracting
particles), we see that the magnitudes of the masses and charges of the charge
carriers are important, that the induced electric fields can be small, and
that energy goes into the mechanical kinetic energy of the charge carriers.
\ However, for large numbers of interacting charges, the mutual interactions
make the magnitudes of the masses and charges unimportant, the induced
electric fields balance the inducing fields, and the energy goes into magnetic
energy of the circuit.

\section{Point-Charge Fields}

\subsection{Point-Charge Fields for General Motion}

Although students are familiar with the Coulomb electric field of a point
charge, many do not study electromagnetism to the point that they see the full
retarded point-charge fields from the Li\'{e}nard-Wiechert
potentials\cite{J664}%
\begin{equation}
\mathbf{E}(\mathbf{r},t)=e\left[  \frac{(1-v_{e}^{2}/c^{2})(\mathbf{n}%
-\mathbf{v}_{e}/c)}{(1-\mathbf{n}\cdot\mathbf{v}_{e}/c)^{3}|\mathbf{r}%
-\mathbf{r}_{e}|^{2}}\right]  _{t_{ret}}+\frac{e}{c^{2}}\left[  \frac
{\mathbf{n}\times\{(\mathbf{n}-\mathbf{v}_{e}/c)\times\mathbf{a}_{e}%
\}}{(1-\mathbf{n}\cdot\mathbf{v}_{e}/c)^{3}|\mathbf{r}-\mathbf{r}_{e}%
|}\right]  _{t_{ret}} \label{E}%
\end{equation}
and
\begin{equation}
\mathbf{B}(\mathbf{r},t)=e\mathbf{n}_{t_{ret}}\times\mathbf{E(r},t) \label{B}%
\end{equation}
where the unit vector $\mathbf{n}=(\mathbf{r-r}_{e})/|\mathbf{r-r}_{e}|,$ and
where the position $\mathbf{r}_{e}(t),$ velocity $\mathbf{v}_{e}(t),~$\ and
acceleration $\mathbf{a}_{e}(t)$ of the charge $e$ must be evaluated at the
retarded time $t_{ret}$ such that $c|t-t_{ret}|=|\mathbf{r}-\mathbf{r}%
_{e}(t_{ret})|.$ \ These field expressions, together with Newton's second law
for the Lorentz force, give the causal interactions between point
charges.\cite{G439} \ The only thing missing from this formulation of
classical electrodynamics is the possible existence of a homogeneous solution
of Maxwell's equations (such as, for example, a plane wave) which might
interact with the charged particles. \ However, electromagnetic induction
fields are distinct from homogeneous radiation fields, and therefore we expect
that the fields of Faraday induction can be treated as having their origin
from charged particle motions. The exploration of this point-charge point of
view in connection with Faraday induction is the subject of the present article.

\subsection{Low-Velocity-Small-Distance Approximation without Retardation\ }

Since the electric field of Faraday induction is distinct from the Coulomb
field of a stationary charge, we expect this induction field to involve the
additional velocity- and acceleration-dependent terms of Eqs. (\ref{E}).
\ \ The use of the full expressions involving retardation in Eqs. (\ref{E})
and (\ref{B}) can be a formidable task. \ For the radiation fields which fall
off as $1/r$ at large distances, the presence of retardation cannot be avoided
as the signal travels from the source charge to the distant field point.
\ However, for field points which are near point charges which are moving at
low velocities, it is possible to derive from Eqs. (\ref{E}) and (\ref{B})
approximate expressions for the electric and magnetic fields which involve no
retardation. \ The task of approximation is not trivial and is carried out in
the textbook \textit{Electrodynamics} by Page and Adams\cite{Electrodynamics}
giving%
\begin{align}
\mathbf{E}_{j}(\mathbf{r,t})  &  =e_{j}\frac{(\mathbf{r}-\mathbf{r}_{j}%
)}{|\mathbf{r}-\mathbf{r}_{j}|^{3}}\left[  1+\frac{\mathbf{v}_{j}^{2}}{2c^{2}%
}-\frac{3}{2}\left(  \frac{\mathbf{v}_{j}\cdot(\mathbf{r}-\mathbf{r}_{j}%
)}{c|\mathbf{r}-\mathbf{r}_{j}|}\right)  ^{2}\right] \nonumber\\
&  -\frac{e_{j}}{2c^{2}}\left(  \frac{\mathbf{a}_{j}}{|\mathbf{r}%
-\mathbf{r}_{j}|}+\frac{\mathbf{a}_{j}\cdot(\mathbf{r}-\mathbf{r}%
_{j})(\mathbf{r}-\mathbf{r}_{j})}{|\mathbf{r}-\mathbf{r}_{j}|^{3}}\right)
+O(1/c^{3}) \label{e6}%
\end{align}
and
\begin{equation}
\mathbf{B}_{j}(\mathbf{r},t)=e_{j}\frac{\mathbf{v}_{j}}{c}\times
\frac{(\mathbf{r}-\mathbf{r}_{j})}{|\mathbf{r}-\mathbf{r}_{j}|^{3}}+O(1/c^{3})
\label{e7}%
\end{equation}
where in Eq. (\ref{e6}) the quantity $\mathbf{a}_{j}$ refers to the
acceleration of the $j$th particle. \ \ These approximate expressions are
power series in $1/c$ where $c$ is the speed of light in vacuum. \ The
approximate fields in Eqs. (\ref{e6}) and (\ref{e7}) can be quite useful; they
were used by Page and Adams\cite{PageAdams} to discuss \textquotedblleft
Action and Reaction Between Moving Charges,\textquotedblright\ and by the
current author when interested in Lorentz-transformation properties of energy
and momentum\cite{B1985} and questions of mass-energy equivalence.\cite{B1998}
\ Here the approximate expressions for the fields are exactly what is needed
to understand Faraday induction from a particle point of view. \ 

The approximate fields in Eqs. (\ref{e6}) and (\ref{e7}) also correspond to
those which arise from the Darwin Lagrangian\cite{Darwin}
\begin{align}
\mathcal{L}  &  \mathcal{=}%
%TCIMACRO{\tsum \limits_{i=1}^{i=N}}%
%BeginExpansion
{\textstyle\sum\limits_{i=1}^{i=N}}
%EndExpansion
m_{i}c^{2}\left(  -1+\frac{\mathbf{v}_{i}^{2}}{2c^{2}}+\frac{(\mathbf{v}%
_{i}^{2})^{2}}{8c^{4}}\right)  -\frac{1}{2}%
%TCIMACRO{\tsum \limits_{i=1}^{i=N}}%
%BeginExpansion
{\textstyle\sum\limits_{i=1}^{i=N}}
%EndExpansion%
%TCIMACRO{\tsum \limits_{j\neq i}}%
%BeginExpansion
{\textstyle\sum\limits_{j\neq i}}
%EndExpansion
\frac{e_{i}e_{j}}{|\mathbf{r}_{i}-\mathbf{r}_{j}|}\nonumber\\
&  +\frac{1}{2}%
%TCIMACRO{\tsum \limits_{i=1}^{i=N}}%
%BeginExpansion
{\textstyle\sum\limits_{i=1}^{i=N}}
%EndExpansion%
%TCIMACRO{\tsum \limits_{j\neq i}}%
%BeginExpansion
{\textstyle\sum\limits_{j\neq i}}
%EndExpansion
\frac{e_{i}e_{j}}{2c^{2}}\left[  \frac{\mathbf{v}_{i}\cdot\mathbf{v}_{j}%
}{|\mathbf{r}_{i}-\mathbf{r}_{j}|}+\frac{\mathbf{v}_{i}\cdot(\mathbf{r}%
_{i}-\mathbf{r}_{j})\mathbf{v}_{j}\cdot(\mathbf{r}_{i}-\mathbf{r}_{j}%
)}{|\mathbf{r}_{i}-\mathbf{r}_{j}|^{3}}\right] \nonumber\\
&  -%
%TCIMACRO{\tsum \limits_{i=1}^{i=N}}%
%BeginExpansion
{\textstyle\sum\limits_{i=1}^{i=N}}
%EndExpansion
e_{i}\Phi_{ext}(\mathbf{r}_{i},t)+%
%TCIMACRO{\tsum \limits_{i=1}^{i=N}}%
%BeginExpansion
{\textstyle\sum\limits_{i=1}^{i=N}}
%EndExpansion
e_{i}\frac{\mathbf{v}_{i}}{c}\cdot\mathbf{A}_{ext}(\mathbf{r}_{i},t)
\label{e4}%
\end{align}
where the last line includes the scalar potential $\Phi_{ext}$ and vector
potential $\mathbf{A}_{ext}$ associated with the external electromagnetic
fields. \ The Darwin Lagrangian omits radiation but expresses accurately the
interaction of charged particles through order $1/c^{2}.$ \ The Darwin
Lagrangian continues to appear in advanced textbooks\cite{Darwin}, but the
approximate expressions (\ref{e6}) and (\ref{e7}) seem to have disappeared
from the consciousness of most contemporary physicists. \ The Lagrangian
equations of motion from the Darwin Lagrangian can be rewritten in the form of
Newton's second law $d\mathbf{p}/dt=d(m\gamma\mathbf{v})/dt=\mathbf{F}$ with
$\gamma=(1-v^{2}/c^{2})^{-1/2}.$ \ In this Newtonian form, we have%
\begin{align}
&  \frac{d}{dt}\left[  \frac{m_{i}\mathbf{v}_{i}}{(1-\mathbf{v}_{i}^{2}%
/c^{2})^{1/2}}\right] \nonumber\\
&  \approx\frac{d}{dt}\left[  m_{i}\left(  1+\frac{\mathbf{v}_{i}^{2}}{2c^{2}%
}\right)  \mathbf{v}_{i}\right]  =e_{i}\mathbf{E}+e_{i}\frac{\mathbf{v}_{i}%
}{c}\times\mathbf{B}\nonumber\\
&  =e_{i}\left(  \mathbf{E}_{ext}(\mathbf{r}_{i},t)+%
%TCIMACRO{\tsum \limits_{j\neq i}}%
%BeginExpansion
{\textstyle\sum\limits_{j\neq i}}
%EndExpansion
\mathbf{E}_{j}(\mathbf{r}_{i},t)\right)  +e_{i}\frac{\mathbf{v}_{i}}{c}%
\times\left(  \mathbf{B}_{ext}(\mathbf{r}_{i},t)+%
%TCIMACRO{\tsum \limits_{j\neq i}}%
%BeginExpansion
{\textstyle\sum\limits_{j\neq i}}
%EndExpansion
\mathbf{B}_{j}(\mathbf{r}_{i},t)\right)  \label{e5}%
\end{align}
with the Lorentz force on the $i$th particle arising from the external
electromagnetic fields and from the electromagnetic fields of the other
particles. \ The electromagnetic fields due to the $j$th particle are given
through order $v^{2}/c^{2}$ by exactly the approximate expressions appearing
in Eqs. (\ref{e6}) and (\ref{e7}).

\section{Electromagnetic Induction}

Electromagnetic induction was discovered by Michael Faraday, not as a
motion-dependent modification of Coulomb's law, but rather in terms of emfs
producing currents in circuits. \ This circuit-based orientation remains the
way that electromagnetic induction is discussed in textbooks today. \ \ Now
the emf in a circuit is the closed line integral around the circuit of the
force per unit charge $\mathbf{f}$ acting on the charges of the circuit, $emf=%
%TCIMACRO{\toint }%
%BeginExpansion
{\textstyle\oint}
%EndExpansion
\mathbf{f}\cdot\mathbf{dr}$. \ Faraday's emfs were associated with changing
magnetic fluxes, and Faraday's law of electromagnetic induction in a circuit
is given by%
\begin{equation}
emf_{F}=-\frac{1}{c}\frac{d\Phi}{dt}%
\end{equation}
where $\Phi$ is the magnetic flux through the circuit. \ 

As correctly emphasized in some textbooks,\cite{G303} electromagnetic
induction in a circuit can arise in two distinct aspects. \ The
\textquotedblleft motional emf\textquotedblright\ in a circuit which is moving
through an unchanging magnetic field can be regarded as arising from the
magnetic Lorentz force acting on the mobile charges of the moving circuit.
\ On the other hand, when the circuit is stationary in space but the current
in the circuit is changing, new electric fields arise in space. \ These new
electric fields can cause an emf in an adjacent circuit (mutual inductance) or
in the original circuit itself (self-inductance); the new electric fields are
precisely those appearing due to the motions of the charges of the circuit as
given in Eq. (\ref{E}), or, through order $v^{2}/c^{2},$ as given in Eq.
(\ref{e6}).\ \ It is these electric fields which are the subject of our
discussion of Faraday induction.

It should be noted that for steady-state currents in a multiparticle circuit
with large numbers of charges where the charge density and current density are
time-independent, all the complicating motion-dependent terms in Eqs.
(\ref{E})-(\ref{B}) or (\ref{e6})-(\ref{e7}) beyond the first leading term in
$1/c$ actually cancel, so that the electromagnetic fields can be calculated
simply using Coulomb's law and the Biot-Savart Law.\cite{J-14} \ However, for
time-varying charge densities and/or current densities, the motion-dependent
terms in Eqs. (\ref{E})-(\ref{B}) or (\ref{e6})-(\ref{e7}) do not cancel and
indeed provide the Faraday induction fields.

If an external emf $emf_{ext}$\ is present in a continuous circuit with a
self-inductance $L$ and resistance $\mathcal{R}$, the current $i$ in the
circuit is given by the differential equation%
\begin{equation}
emf_{ext}=L\frac{di}{dt}+i\mathcal{R} \label{e1}%
\end{equation}
where the term $Ldi/dt$ corresponds to the negative of the Faraday-induced emf
associated with the changing current in the circuit. Here in traditional
electromagnetic theory, the self-inductance $L$ of a rigid circuit is a
time-independent quantity which depends only upon the geometry of the
circuit.\ The energy balance for the circuit is found by multiplying Eq.
(\ref{e1}) by the current $i$%
\begin{equation}
emf_{ext}\times i=\frac{d}{dt}\left(  \frac{1}{2}Li^{2}\right)  +i^{2}%
\mathcal{R} \label{e2}%
\end{equation}
corresponding to a power $emf_{ext}\times i$ delivered by the external emf
going into the time-rate-of-change of magnetic energy $(1/2)Li^{2}$ stored in
the inductor and the power $i^{2}\mathcal{R}$ lost in the resistor.

Although the energy analysis for Eq. (\ref{e2}) seems natural, the
differential equation (\ref{e1}) presents some unusual aspects if we consider
the circuit from the particle point of view. \ If at time $t=0$ the constant
external emf $emf_{F}$ is applied to the circuit and the current is zero,
$i(0)=0,$ then the Faraday induced emf, $emf_{F}=-Ldi/dt,$ must exactly cancel
the external emf $emf_{ext}\,\ $so that $emf_{ext}-Ldi/dt=0$ at time $t=0.$
\ Phrased in terms of forces per unit charge applied to the circuit, the
Faraday induced electric field $\mathbf{E}_{F}$ must exactly cancel the
external force per unit charge $\mathbf{f}_{ext}$ associated with the external
emf. \ Indeed, if the resistance $\mathcal{R}$ of the circuit becomes
vanishingly small, $\mathcal{R}\rightarrow0,$ then this canceling balance of
the Faraday electric field $\mathbf{E}_{F}$ against the force per unit charge
$\mathbf{f}_{ext}$ associated with the external emf holds at all times, and
yet the current increases at a constant rate following $di/dt=emf_{ext}/L.$
\ But if the net force per unit charge goes to zero, why do the charges
accelerate so as to produce a changing current $di/dt?$ \ After all, in
classical mechanics, $\mathbf{F}_{R}=m\mathbf{a};$ it is the resultant force
$\mathbf{F}_{R}$ on a particle which determines the acceleration $\mathbf{a}$
of the particle of mass $m.$ \ Thus we expect that if the resultant force on a
particle of mass $m$ is zero, then the mass does not accelerate. \ However,
electromagnetism involves some aspects which are different from what is
familiar in nonrelativistic mechanics, and electromagnetic circuit theory
involves some unmentioned approximations. \ In this article, we wish to
explore these differences and unmentioned approximations by using a particle
model in connection with Faraday induction. \ We will note the approximations
involved in Eq. (\ref{e1}) which lead to the troubling apparent contradiction
with Newton's second law. \ 

\section{Detailed Discussion of Faraday Induction in a Simple Hypothetical
Circuit}

\subsection{Model for a Detailed Discussion}

Here we would like to explore Faraday induction in some detail for the
simplest possible circuit in hopes of obtaining some physical insight.
\ Accordingly, we will discuss a hypothetical circuit consisting of $N$
equally-spaced particles of mass $m$ and charge $e$ which are constrained by
centripetal forces of constraint to move in a circular orbit of radius $R$ in
the $xy$-plane centered on the origin. \ A balancing negative charge to give
the circuit neutrality can be thought of as a uniform line charge in the orbit
or else as a single compensating charge at the center of the orbit. \ The
choice does not influence the analysis to follow. \ The system may be thought
of as consisting of charged beads sliding on a frictionless ring. \ There is
no frictional force and hence no resistance $\mathcal{R}$ in the model. \ The
model is intended as a rough approximation to a circular loop of wire of small
cross-section.\cite{Essen2} \ \ 

We now imagine that a constant external force per unit charge $\mathbf{f}%
_{ext}$ is applied in a circular pattern in the tangential $\widehat{\phi}%
$-direction, $\mathbf{f}_{ext}=\widehat{\phi}f_{ext},$ to all the charges of
the ring. \ One need not specify the source of this external force per unit
charge $\mathbf{f}_{ext}$, but one example of such a situation involves an
axially-symmetric magnetic field applied perpendicular to the plane of the
circular orbit in the $-\widehat{z}$-direction which is increasing in
magnitude at a constant rate. The external emf around the circular orbit is
given by
\begin{equation}
emf_{ext}=%
%TCIMACRO{\toint }%
%BeginExpansion
{\textstyle\oint}
%EndExpansion
\mathbf{f}_{ext}\cdot\mathbf{dr}=2\pi Rf_{ext}\text{ .} \label{e3}%
\end{equation}
The external force per unit charge $\mathbf{f}_{ext}$ places a tangential
force $\mathbf{F}_{i}=e_{i}\widehat{\phi}_{i}f_{ext}$ on the $i$th particle
located at $\mathbf{r}_{i}.$ \ The Faraday inductance of the charged-particle
system is determined by the response of all the particles $e_{i}$ in the
circular orbit. \ 

\subsection{One-Particle Model for a Circuit}

\subsubsection{Motion of the Charged Particle}

We start with the case when there is only one charged particle of mass $m$ and
charge $e$ in the circular orbit. \ In this case, the tangential acceleration
$a_{\phi}$ of the charged particle $e$ in the circular orbit arises from the
(tangential) force of only the external force per unit charge $\mathbf{f}%
_{ext},$ since the centripetal forces of constraint are all radial forces.
\ From Eq. (\ref{e5}), written for a single particle and with $d(m\gamma
v)/dt=m\gamma^{3}a_{\phi}$ where $\gamma=(1-v^{2}/c^{2})^{-1/2}$, we have
\begin{equation}
a_{\phi}=\frac{ef_{ext}}{m\gamma^{3}} \label{e8}%
\end{equation}
where $f_{ext}$ is the magnitude of the tangential force per unit charge due
to the external emf $emf_{ext}$ at the position of the charge $e.$

\subsubsection{Magnetic Field of the Charged Particle \ }

The magnetic field $\mathbf{B}_{e}$ at the center of the circular orbit due to
the accelerating charge $e$ is given by Eq. (\ref{e7})
\begin{equation}
\mathbf{B}_{e}(0,t)=\widehat{k}e\frac{v}{cR^{2}} \label{e9}%
\end{equation}
where the velocity $v$ is increasing since the external force per unit charge
$\mathbf{f}_{ext}$ gives a positive charge $e$ a positive acceleration in the
$\widehat{\phi}$-direction. This magnetic field $\mathbf{B}_{e}$ produced by
the orbiting charge $e$ is increasing in the $\widehat{z}$-direction, which is
in the opposite direction from the increasing external magnetic field which
could have created the external force per unit charge $\mathbf{f}_{ext}$ and
the external emf $emf_{ext}$ in Eq. (\ref{e3}). \ 

\subsubsection{Induced Electric Field from Faraday's Law}

Associated with this changing magnetic field $\mathbf{B}_{e}$ created by the
orbiting charge $e,$ there should be an induced electric field $\mathbf{E}%
_{e}(\mathbf{r},t)$ according to Faraday's law. \ Thus averaging over the
circular motion of the charge $e$, we expect an average induced tangential
electric field $\left\langle E_{e\phi}(r)\right\rangle $ at a distance $r$
from the center of the circular orbit (where $r<<R$ so that the magnetic field
$\mathbf{B}_{e}$ has approximately the value $\mathbf{B}(0,t)$ at the center)
given from Eq. (\ref{e9}) by
\begin{align}
2\pi r\,\left\langle E_{e\phi}(r)\right\rangle  &  =emf_{e}=-\frac{1}{c}%
\frac{d\Phi_{e}}{dt}=-\frac{1}{c}\frac{d}{dt}[B_{e}(0,t)\pi r^{2}]\nonumber\\
&  ==-\frac{1}{c}\frac{d}{dt}\left[  e\frac{v}{cR^{2}}\pi r^{2}\right]
=-\frac{1}{c}\left(  e\frac{a_{\phi}}{cR^{2}}\right)  \pi r^{2} \label{eee10}%
\end{align}
since $dv/dt=a_{\phi.}$ \ Using Eq. (\ref{e8}), the average tangential
electric field follows from Eq. (\ref{eee10}) as%
\begin{equation}
\left\langle \mathbf{E}_{e\phi}(r,t)\right\rangle =-\widehat{\phi}\frac
{e^{2}rf_{ext}}{2mc^{2}\gamma^{3}R^{2}} \label{e11}%
\end{equation}
This derivation of the Faraday induction field corresponds to that familiar
from the textbook approach.

\subsubsection{Induced Electric Field from the Approximate Point-Charge
Fields}

We will now show that this induced average tangential electric field
$\left\langle \mathbf{E}_{e\phi}(r,t)\right\rangle $ is exactly the average
electric field due to the charge $e$ obtained by use of the approximate
electric field expression given in Eq. (\ref{e6}). \ Thus we assume that the
charge $e$ is located momentarily at $\mathbf{r}_{e}=\widehat{x}R\cos\phi
_{e}+\widehat{y}R\sin\phi_{e},$ and we average the electric field
$\mathbf{E}_{e}(\mathbf{r},t)$ due to $e$ over the phase $\phi_{e}$. \ Since
the entire situation is axially symmetric when averaged over $\phi_{e},$ we
may take the field point along the $x$-axis at $\mathbf{r}=\widehat{x}r$, and
later generalize to cylindrical coordinates. \ The velocity fields given in
the first line of Eq. (\ref{e6}) point from the charge $e$ to the field point.
\ Also, the velocity fields are even if the sign of the velocity
$\mathbf{v}_{e}$ is changed to $-\mathbf{v}_{e}.$ \ Thus the velocity fields
when averaged over the circular orbit can point only in the radial direction.
\ The acceleration fields arising from the centripetal acceleration of the
charge will also point in the radial direction. \ Since we are interested in
the average tangential component of the field $\mathbf{E}_{e}$, we need to
average over only the tangential acceleration terms in the second line of Eq.
(\ref{e6}). \ If the field point is close to the center of the circular orbit
so that $r<<R,$ then we may expand in powers of $r/R;$ we retain only the
first-order terms, giving $|\widehat{x}r-\mathbf{r}_{e}|^{-1}\approx
R^{-1}(1+\widehat{x}r\cdot\mathbf{r}_{e}/R^{2})$ and $|\widehat{x}%
r-\mathbf{r}_{e}|^{-3}\approx R^{-3}(1+3\widehat{x}r\cdot\mathbf{r}_{e}%
/R^{2}).$ \ Then the average tangential component of the electric field due to
the charge $e$ can be written as
\begin{align}
\left\langle \mathbf{E}_{e\phi}(\widehat{x}r,t)\right\rangle  &  =\left\langle
-\frac{e}{2c^{2}}\left(  \frac{\mathbf{a}_{e\phi}}{|\widehat{x}r-\mathbf{r}%
_{e}|}+\frac{\mathbf{a}_{e\phi}\cdot(\widehat{x}r-\mathbf{r}_{e}%
)(\widehat{x}r-\mathbf{r}_{e})}{|\widehat{x}r-\mathbf{r}_{e}|^{3}}\right)
\right\rangle \nonumber\\
&  =\left\langle -\frac{e}{2c^{2}}\left[  \frac{\mathbf{a}_{e\phi}}{R}\left(
1+\frac{\widehat{x}r\cdot\mathbf{r}_{e}}{R^{2}}\right)  +\frac{\mathbf{a}%
_{e\phi}\cdot(\widehat{x}r-\mathbf{r}_{e})(\widehat{x}r-\mathbf{r}_{e})}%
{R^{3}}\left(  1+\frac{3\widehat{x}r\cdot\mathbf{r}_{e}}{R^{2}}\right)
\right]  \right\rangle \text{ .} \label{e12}%
\end{align}
Now we average over the phase $\phi_{e}$ with $\mathbf{r}_{e}=\widehat{x}%
R\cos\phi_{e}+\widehat{y}R\sin\phi_{e}$ and $\mathbf{a}_{e\phi}=a_{e\phi
}(\mathbf{-}\widehat{x}\sin\phi_{e}+\widehat{y}\cos\phi_{e}).$ \ We note that
$\left\langle \mathbf{a}_{e\phi}\right\rangle =0,$ $\mathbf{a}_{e\phi}%
\cdot\mathbf{r}_{e}=0,$ $\left\langle \mathbf{a}_{e\phi}(\widehat{x}%
\cdot\mathbf{r}_{e})\right\rangle =\widehat{y}a_{e\phi}R/2=-\left\langle
(\mathbf{a}_{e\phi}\cdot\widehat{x})\mathbf{r}_{e}\right\rangle .$ \ After
averaging and retaining terms through order $r/R $, equation (\ref{e12})
becomes%
\begin{equation}
\left\langle \mathbf{E}_{e\phi}(\widehat{i}r,t)\right\rangle =-\widehat{y}%
\frac{ea_{\phi}r}{2c^{2}R^{2}} \label{e13}%
\end{equation}
which is in agreement with our earlier results in Eqs. (\ref{eee10}) and
(\ref{e11}). \ Thus indeed the electric field of Faraday induction in this
case arises from the acceleration of the charged current carrier of the circuit.

\subsubsection{Limit on the Induced Electric Field}

We are now in a position to comment on the average response of our
one-particle circuit to the applied external emf. If the source of the
external $emf_{ext}$ is a changing magnetic field, then this situation
corresponds to the traditional example for diamagnetism within classical
electromagnetism.\cite{diamag} \ For this one-particle example, the response
depends crucially upon the mass $m$ and charge $e$ of the particle. \ When the
mass $m$ is large, the acceleration of the charge is small; therefore the
induced tangential electric field $\mathbf{E}_{e\phi}$ in Eq. (\ref{e11}) is
small. \ This large-mass situation is what is usually assumed in examples of
charged rings responding to external emfs.\cite{Feynman} \ On the other hand,
if we try to increase the induced electromagnetic field $\mathbf{E}_{e\phi}$
by making the mass $m$ small, we encounter a fundamental limit of
electromagnetic theory. \ The allowed mass $m$ is limited below by
considerations involving the classical radius of the electron $r_{cl}%
=e^{2}/(mc^{2}).$ \ Classical electromagnetic theory is valid only for
distances large compared to the classical radius of the electron. \ Thus in
our example where the radius $R$ of the orbit is a crucial parameter, we must
have $R>>r_{cl}.$ This means we require the mass $m>>e^{2}/(Rc^{2})$ and so
$e^{2}/(mc^{2}R)<<1.$ \ Combining this limit with $r/R<1,$ and $1<\gamma$
leads to a limit on the magnitude of the induced electric field in Eq.
(\ref{e11})
\begin{equation}
\left\langle E_{e\phi}(r,t)\right\rangle <<f_{ext}\text{ \ \ for
\ \ }r<R\text{ .} \label{e14}%
\end{equation}
The induced electric field of a one-particle circuit is small compared to the
external force per unit charge associated with the external emf. \ 

\subsubsection{Energy Balance}

We also note that the power delivered by the external force per unit charge
goes into kinetic energy of the orbiting particle. \ Thus if take the
Newton's-second-law equation giving Eq. (\ref{e8}) and multiply by the speed
$v$ of the particle, we have%
\begin{equation}
\frac{d}{dt}(m\gamma v)v=\frac{d}{dt}(m\gamma c^{2})=m\gamma^{3}a_{\phi
}v=ef_{ext}v \label{e15}%
\end{equation}
so that the power $ef_{ext}v$ delivered to the charge $e$ by the external
force goes into kinetic energy of the particle. \ 

The situation of a one-particle circuit can be summarized as follows. \ For
the one-particle circuit, the induced electric field is small compared to the
external electric field and depends explicitly upon the particle's mass and
charge, while the energy transferred by the external field goes into kinetic
energy of the one charged particle. \ Clearly this is not the situation which
we usually associate with electromagnetic induction for circuit problems.

\subsection{Multiparticle Model for a Circuit}

\subsubsection{Motion of the Charged Particles}

In order to make contact with the usual discussion of Faraday inductance in a
circuit, we must go to the situation of many particles, each one of charge $e$
and mass $m$. \ However, if we take the one-particle circuit above and simply
superimpose the fields corresponding to $N$ equally-spaced charges while
maintaining the acceleration appropriate for the single-particle case, we
arrive at a completely false result. \ Thus if we take Eq. (\ref{e11}) for the
Faraday-induced average electric field $\left\langle \mathbf{E}_{e\phi
}(r,t)\right\rangle $\ due to a single particle of charge $e$ and mass $m$
accelerating in the external force per unit charge $f_{ext}$ and then simply
multiply by the number of charges, we have a result which is linear in $N$ and
increases without bound. \ Thus merely extrapolating the one-particle circuit
suggests that the Faraday induced electric field arising from many charges
might far exceed the inducing force per unit charge $f_{ext}$. \ In order to
obtain a correct understanding of the physics, we must include the mutual
interactions between the accelerating charges of the circuit. \ Now with these
mutual interactions, the force on any charge in the circular orbit is not just
the external force $ef_{ext}$ due to the original external electric field, and
the acceleration of any charge is not given by $a_{\phi}=ef_{ext}/(m\gamma
^{3})$. \ Now the force on any charge is a sum of the force due to the
original external force per unit charge plus the forces due to the fields of
all the other charged particles in the circular orbit as given in Eq.
(\ref{e5}). \ The magnetic force $e_{i}\mathbf{v}_{i}\times\mathbf{B}/c$ is
simply a deflection and does not contribute to the the tangential acceleration
of the charge $e_{i}$. \ Thus the equation of motion for the $i$th particle
becomes%
\begin{align}
&  \frac{d}{dt}(m_{i}\gamma_{i}\mathbf{v}_{i})\cdot\widehat{\phi}\nonumber\\
&  =m_{i}\gamma_{i}^{3}a_{i\phi}=m_{i}\gamma_{i}^{3}R\frac{d^{2}\phi_{i}%
}{dt^{2}}\nonumber\\
&  =\widehat{\phi}_{i}\cdot e_{i}\left\{  \mathbf{f}_{ext}\mathbf{(r}%
_{i})+\sum\limits_{j\neq i}e_{j}\frac{(\mathbf{r}_{i}-\mathbf{r}_{j}%
)}{|\mathbf{r}_{i}-\mathbf{r}_{j}|^{3}}-\sum\limits_{j\neq i}\frac{e_{j}%
}{2c^{2}}\left(  \frac{\mathbf{a}_{j}}{|\mathbf{r}_{i}-\mathbf{r}_{j}|}%
+\frac{\mathbf{a}_{j}\cdot(\mathbf{r}_{i}-\mathbf{r}_{j})(\mathbf{r}%
_{i}-\mathbf{r}_{j})}{|\mathbf{r}_{i}-\mathbf{r}_{j}|^{3}}\right)  \right\}
\label{ee16}%
\end{align}
where it is understood that the factor $\gamma_{i}=(1-v_{i}^{2}/c^{2})^{-1/2}$
should be expanded through second order in $v_{i}/c$ so as to be consistent
with the remaining terms arising from the approximate field expression
(\ref{e6}). \ Since the particles are equally spaced around the circular orbit
and all have the same charge $e$ and mass $m $, the situation is axially
symmetric. \ The equation of motion for every charge takes the same form, and
the angular acceleration of each charge is the same, $d^{2}\phi_{i}%
/dt^{2}=d^{2}\phi/dt^{2}.$ \ For simplicity of calculation, we will take the
$N$th particle along the $x$-axis so that $\phi_{N}=0,$ $\mathbf{r}%
_{N}=\widehat{x}R,$ and $\widehat{\phi}_{N}=\widehat{y}.$ \ The other
particles are located at $\mathbf{r}_{j}=\widehat{x}R\cos(2\pi
j/N)+\widehat{y}R\sin(2\pi j/N),$ corresponding to an angle $\phi_{j}=2\pi
j/N$ for $j=1,2,...,N-1.\ $The tangential acceleration of the $j$th particle
is given by $\mathbf{a}_{j\phi}=(d^{2}\phi/dt^{2})[-\widehat{x}R\sin(2\pi
j/N)+\widehat{y}R\cos(2\pi j/N)].$ \ By symmetry, it is clear that the
electrostatic fields, the velocity fields, and the centripetal acceleration
fields of the other particles can not contribute to the tangential electric
field at particle $N.$ \ The equation of motion for the tangential
acceleration for each charge in the circular orbit is the same as that for the
$N$th particle, which from Eq. (\ref{ee16}) is
\begin{equation}
m\gamma^{3}R\frac{d^{2}\phi}{dt^{2}}=\left\{  ef_{ext}-\sum\limits_{j=1}%
^{j=N-1}\frac{e^{2}}{2c^{2}}\left(  \frac{\widehat{y}\cdot\mathbf{a}_{j\phi}%
}{|\widehat{x}R-\mathbf{r}_{j}|}+\frac{\mathbf{a}_{j\phi}\cdot(\widehat{x}%
R-\mathbf{r}_{j})\widehat{y}\cdot(\widehat{x}R-\mathbf{r}_{j})}{|\widehat{x}%
R-\mathbf{r}_{j}|^{3}}\right)  \right\}  \text{ .} \label{e16a}%
\end{equation}
Now we evaluate the distance between the $j$th particle and the $N$th particle
in the circular orbit as
\begin{equation}
|\widehat{x}R-\mathbf{r}_{j}|=[2R^{2}-2R^{2}\cos(2\pi j/N)]^{1/2}=[4R^{2}%
\sin^{2}(\pi j/N)]^{1/2}=|2R\sin(\pi j/N)| \label{e17a}%
\end{equation}
while
\begin{equation}
\widehat{y}\cdot\mathbf{a}_{j\phi}=(d^{2}\phi/dt^{2})R\cos(2\pi j/N)
\label{e17b}%
\end{equation}
and
\begin{equation}
\mathbf{a}_{j}\cdot(\widehat{x}R-\mathbf{r}_{j})\widehat{y}\cdot
(\widehat{x}R-\mathbf{r}_{j})=(d^{2}\phi/dt^{2})R[-R\sin(2\pi j/N)][-R\sin
(2\pi j/N)]\text{ .} \label{e17c}%
\end{equation}
\ Then equation (\ref{e16a}) becomes%
\begin{align}
&  m\gamma^{3}R\frac{d^{2}\phi}{dt^{2}}\nonumber\\
&  =ef_{ext}-\frac{d^{2}\phi}{dt^{2}}\sum\limits_{j=1}^{j=N-1}\frac{e^{2}%
}{2c^{2}}\left(  \frac{R\cos(2\pi j/N)}{|2R\sin(\pi j/N)|}+\frac{R[R\sin(2\pi
j/N)][R\sin(2\pi j/N)]}{|2R\sin(\pi j/N)|^{3}}\right) \nonumber\\
&  =ef_{ext}-\frac{d^{2}\phi}{dt^{2}}\sum\limits_{j=1}^{j=N-1}\frac{e^{2}%
}{2c^{2}}\left(  \frac{2-3\sin^{2}(\pi j/N)}{2\sin(\pi j/N)}\right)
\label{e18}%
\end{align}
or, solving for $d^{2}\phi/dt^{2},$%
\begin{equation}
\frac{d^{2}\phi}{dt^{2}}=ef_{ext}\left[  m\gamma^{3}R+\sum\limits_{j=1}%
^{j=N-1}\frac{e^{2}}{2c^{2}}\left(  \frac{2-3\sin^{2}(\pi j/N)}{2\sin(\pi
j/N)}\right)  \right]  ^{-1}\text{ .} \label{e19}%
\end{equation}
If there is only one particle on the frictionless ring so that $N=1$, the sum
disappears, and the tangential acceleration corresponds to the result obtained
earlier in Eq. (\ref{e8}) above with $Rd^{2}\phi/dt^{2}=a_{\phi}$. \ We note
that the mass term in Eq. (\ref{e19}) remains unchanged by the number of
particles while the sum increases with each additional particle. \ Thus if
there are many particles, then the electric field at particle $i$ due to the
other particles $j$\ can lead to so large a sum in Eq. (\ref{e19}) that the
mass contribution $m\gamma^{3}R$ becomes insignificant. \ In this case, the
common angular acceleration of each particle becomes from Eq. (\ref{e19})%
\begin{equation}
\frac{d^{2}\phi}{dt^{2}}\approx\frac{2c^{2}}{e}f_{ext}\left[  \sum
\limits_{j=1}^{j=N-1}\left(  \frac{2-3\sin^{2}(\pi j/N)}{2\sin(\pi
j/N)}\right)  \right]  ^{-1} \label{e20}%
\end{equation}
\ We see that in this multiparticle situation the angular acceleration no
longer depends upon the mass $m$ of the charge carriers. \ This is the
situation envisioned in the usual textbook treatment of Faraday induction.

\subsubsection{Induced Electric Field}

Furthermore, in this multiparticle situation where the particle mass becomes
insignificant, the left-hand side of Eq. (\ref{e16a}) is negligible, so that
the sum $%
%TCIMACRO{\tsum _{j}}%
%BeginExpansion
{\textstyle\sum_{j}}
%EndExpansion
\mathbf{E}_{ej}(\mathbf{r}_{i})~$of the acceleration fields of all the other
charges $e_{j}$ cancels the external force per unit charge $\mathbf{f}%
_{ext}(\mathbf{r}_{i})$ of the external emf at the position $\mathbf{r}_{i}%
$\ of each charge in the circular orbit%

\begin{equation}
-\mathbf{f}_{ext}\mathbf{(r}_{i}\mathbf{)\approx E}_{e}(\mathbf{r}_{i})=%
%TCIMACRO{\tsum _{j\neq i}}%
%BeginExpansion
{\textstyle\sum_{j\neq i}}
%EndExpansion
\mathbf{E}_{ej}(\mathbf{r}_{i})\text{ .} \label{e21}%
\end{equation}
This situation is analogous to that in electrostatics where the fields of the
charges in a conductor move to new positions so as to cancel the external
force per unit charge at the position of each charge; here the charges
accelerate so that the acceleration fields cancel the external force per unit
charge at the position of each charge. \ Now the induced electric field
$\mathbf{E}_{e}(\mathbf{r})=%
%TCIMACRO{\tsum _{i}}%
%BeginExpansion
{\textstyle\sum_{i}}
%EndExpansion
\mathbf{E}_{ei}(\mathbf{r})$ at a general field point due to the orbit
particles is independent of the charge $e$ of the charge carriers, since the
angular acceleration in Eq. (\ref{e20}) depends inversely on the charge $e,$
and this inverse dependence upon $e$ cancels with the $e$ appearing in Eq.
(\ref{e6}) so as to give an induced electric field which is independent of the
charge on the charge carriers. \ Again, this situation corresponds to that
treated in the textbooks when there is no resistance in the circuit; if there
is no resistance, the self-induced emf cancels the external emf and the total
force per unit charge is zero. \ 

Most physicists find surprising this idea that the net force on each particle
is zero and yet the particles are accelerating. \ Indeed, the zero-force
result from Eq. (\ref{e21}) is only an approximation; the full equation is
given in Eq. (\ref{e18}) where indeed Newton's second law holds. \ There is a
self-consistent relation between the\ common acceleration of each charge and
the back electric force on each charge due to the accelerations of the other
charges. \ In the limit of a large number of charges, we can compute the
common acceleration simply by insisting that the acceleration field back at
any charge due to the acceleration of the other charges should (approximately)
balance the external field. \ The situation is perhaps easier to grasp if we
imagine multiplying Eq. (\ref{e18}) by the common velocity of each particle.
\ Then the equation states that the rate-of-change of a particle's kinetic
energy equals the difference between the power delivered to the particle by
the external force and the power absorbed from the particle as magnetic field
energy. \ Since the acceleration of the particle is very small, the rate of
change of the particle kinetic energy is very small and most of the energy
delivered by the external force goes into magnetic field energy. \ If one
neglects the very small amount of energy going into particle kinetic energy,
then one simply states that the energy delivered by the external force goes
into magnetic field energy. \ This last statement is the approximation which
appears in the textbooks of electromagnetism.

In nonrelativistic classical mechanics (which forms the physical ideas of most
physicists), we expect to be able to store potential energy of relative
position; however, there is no such thing as potential energy of velocity.
\ In classical mechanics, the energy associated with velocity is always
mechanical kinetic energy based on mass times velocity squared. \ In complete
contrast to this situation, electromagnetic theory contains magnetic energy
associated with the velocity of charges and yet not associated with particle
mass. \ For consistency, the electromagnetic theory requires that accelerating
charges cause fields which produce forces on other charges. \ The mutual
interaction of the charges through the Faraday induction fields assures that
the magnetic energy stored indeed required work by some external forces. \ For
a multiparticle system, the net force on each particle may become tiny
compared to the external force on each charge because of the acceleration
fields due to the other charges. \ The charges have a tiny acceleration and
gain a tiny amount of kinetic energy while the work done by the external force
goes into the large amount of energy stored in the magnetic field.

\subsubsection{Self-Inductance of the Circuit}

The value for the self-inductance of the multiparticle circuit can be obtained
from Eq. (\ref{e1}) when the circuit resistance vanishes. If the resistance
vanishes, the external emf $emf_{ext}=2\pi Rf_{ext}$ around the ring equals
the self-inductance $L$ multiplied by the time-rate-of-change of the current
$i=Ne(d\phi/dt)/(2\pi)$%
\begin{equation}
emf_{ext}=2\pi RE_{ext}(R)=L\frac{di}{dt}=L\left(  \frac{Ne}{2\pi}\frac
{d^{2}\phi}{dt^{2}}\right)  \text{ .} \label{e22}%
\end{equation}
Now traditional classical electromagnetism ignores the mass contribution to
the inertia which appears in Eq. (\ref{e19}) and regards the self-inductance
as a geometrical quantity associated with the approximation given in Eq,
(\ref{e20}). \ Thus the self-inductance $L$\ of the multi-charge ring system
from Eq. (\ref{e20}) is%
\begin{equation}
L=\frac{2\pi RE_{e}(R)}{[Ne(d^{2}\phi/dt^{2})/(2\pi)]}=\frac{\left(
2\pi\right)  ^{2}R}{2c^{2}N}\sum\limits_{j=1}^{j=N-1}\left(  \frac{2-3\sin
^{2}(\pi j/N)}{2\sin(\pi j/N)}\right)  \text{ .} \label{e23}%
\end{equation}
We see that the self-inductance of this multiparticle, circular-orbit circuit
is now independent of the mass $m$ or the charge $e$ of the current
carriers.\cite{carrierN} \ As with all expressions for the self-inductance $L
$ of a circuit, here Eq. (\ref{e23}) has dimensions of length divided by
$c^{2}.$ \ In the appendix, the self-inductance $L$ in Eq. (\ref{e23}) for our
hypothetical circuit is connected to the self-inductance of a circular wire of
finite cross-section.

\subsubsection{Energy of the Current Carriers}

The change in energy $\Delta U$ of the system associated with non-zero
velocity $v$ for the charges of the circuit includes both mechanical kinetic
energy $\Delta U_{mechanical}$ and magnetic field energy $\Delta U_{mag}$%
\begin{equation}
\Delta U=\Delta U_{mechanical}+\Delta U_{mag}=%
%TCIMACRO{\dsum \limits_{i=1}^{N}}%
%BeginExpansion
{\displaystyle\sum\limits_{i=1}^{N}}
%EndExpansion
m_{i}c^{2}(\gamma_{i}-1)+\frac{1}{8\pi}%
%TCIMACRO{\dint }%
%BeginExpansion
{\displaystyle\int}
%EndExpansion
d^{3}r\mathbf{B}^{2}\text{ .} \label{DU}%
\end{equation}
The self-inductance $L$ of the circuit is associated solely with the magnetic
energy stored in the circuit $\Delta U_{mag}=(1/2)Li^{2}.$ The magnetic energy
$U_{mag}=$ $%
%TCIMACRO{\tint }%
%BeginExpansion
{\textstyle\int}
%EndExpansion
d^{3}r\mathbf{B}^{2}/(8\pi)$ stored in the circular-orbit circuit is given by
the cross-terms (but not the self-terms) when the magnetic field is squared,
and corresponds to the velocity-dependent double sum in the Darwin Lagrangian
Eq. (\ref{e4}). \ Thus we have%
\begin{align}
\Delta U_{mag}  &  =\frac{1}{2}%
%TCIMACRO{\tsum \limits_{i=1}^{N}}%
%BeginExpansion
{\textstyle\sum\limits_{i=1}^{N}}
%EndExpansion%
%TCIMACRO{\tsum \limits_{j\neq i}}%
%BeginExpansion
{\textstyle\sum\limits_{j\neq i}}
%EndExpansion
\frac{1}{8\pi}%
%TCIMACRO{\tint }%
%BeginExpansion
{\textstyle\int}
%EndExpansion
d^{3}r\,2\mathbf{B}_{ei}(\mathbf{r})\cdot\mathbf{B}_{ej}(\mathbf{r}%
)\nonumber\\
&  =\frac{1}{2}%
%TCIMACRO{\tsum \limits_{i=1}^{N}}%
%BeginExpansion
{\textstyle\sum\limits_{i=1}^{N}}
%EndExpansion%
%TCIMACRO{\tsum \limits_{j\neq i}}%
%BeginExpansion
{\textstyle\sum\limits_{j\neq i}}
%EndExpansion
\frac{e^{2}}{2c^{2}}\left(  \frac{\mathbf{v}_{i}\cdot\mathbf{v}_{j}%
}{|\mathbf{r}_{i}-\mathbf{r}_{j}|}+\frac{\mathbf{v}_{i}\cdot(\mathbf{r}%
_{i}-\mathbf{r}_{j})\mathbf{v}_{j}\cdot(\mathbf{r}_{i}-\mathbf{r}_{j}%
)}{|\mathbf{r}_{i}-\mathbf{r}_{j}|^{3}}\right) \nonumber\\
&  =N%
%TCIMACRO{\tsum \limits_{j=1}^{N-1}}%
%BeginExpansion
{\textstyle\sum\limits_{j=1}^{N-1}}
%EndExpansion
\frac{e^{2}}{2c^{2}}\left(  \frac{v\widehat{y}\cdot\mathbf{v}_{j}}{\left\vert
\widehat{x}R-\mathbf{r}_{j}\right\vert }+\frac{v\widehat{y}\cdot
(-\mathbf{r}_{j})\,\mathbf{v}_{j}\cdot(\widehat{x}R)}{\left\vert
\widehat{x}R-\mathbf{r}_{j}\right\vert ^{3}}\right)  \label{ee24}%
\end{align}
where in the last line of Eq. (\ref{ee24}) we have used $\mathbf{v}_{i}%
\cdot\mathbf{r}_{i}=0$ and have taken advantage of the symmetry to evaluate
the magnetic energy when the $N$th particle is located on the $x$-axis at
$\mathbf{r}_{N}=\widehat{x}R$ and is moving with velocity $\mathbf{v}%
_{N}=\widehat{y}v=\widehat{y}Rd\phi/dt.$ The $j$th particle is located at
$\mathbf{r}_{j}=R[\widehat{x}\cos(2\pi j/N)+\widehat{y}\sin(2\pi j/N)]$ with
velocity $\mathbf{v}_{j}=R(d\phi/dt)[-\widehat{x}\sin(2\pi j/N)+\widehat{y}%
\cos(2\pi j/N)].$ \ Introducing these expressions along with the distance
given in Eq. (\ref{e17a}), the magnetic energy of Eq. (\ref{ee24}) is%
\begin{align}
\Delta U_{mag}  &  =N%
%TCIMACRO{\tsum \limits_{j=1}^{N-1}}%
%BeginExpansion
{\textstyle\sum\limits_{j=1}^{N-1}}
%EndExpansion
\frac{e^{2}}{2c^{2}}\left(  R\frac{d\phi}{dt}\right)  ^{2}\left(  \frac
{\cos(2\pi j/N)}{|2\sin(\pi j/N)|}+\frac{\sin^{2}(2\pi j/N)}{|2\sin(\pi
j/N)|^{3}}\right) \nonumber\\
&  =\frac{1}{2}\left[  \frac{\left(  2\pi\right)  ^{2}R}{2c^{2}N}%
\sum\limits_{j=1}^{j=N-1}\left(  \frac{2-3\sin^{2}(\pi j/N)}{2\sin(\pi
j/N)}\right)  \right]  \left(  \frac{eN}{2\pi}\frac{d\phi}{dt}\right)
^{2}\text{ .} \label{e25}%
\end{align}
\qquad We recognize the current $i=eN(d\phi/dt)/(2\pi)$ and so can read off
the self-inductance of the circuit from $\Delta U_{mag}=(1/2)Li^{2}.$ \ The
expression for the self-inductance $L$ is the same as in Eq. (\ref{e23}).
\ Again for an interacting multiparticle system, the mechanical kinetic energy
increases linearly with the number of particles $N$ while the magnetic energy
increases as $N^{2}.$ In this multiparticle limit, the power $P_{i}%
=ef_{ext}v_{i}$ delivered to the $i$th charge by the external force per unit
charge $\mathbf{f}_{ext}$ associated with the original external emf goes only
slightly into particle kinetic energy but mainly is converted into the
magnetic energy associated with the motion of the charges on the ring. \ This
stored magnetic energy is exactly what is given by the expression
$(1/2)Li^{2}.$ \ 

For our example of an external emf acting on a circular charged-particle
circuit, the situation with many particles is totally transformed from the
situation with only one particle. \ As charged particles of mass $m$ and
charge $e$ are added to the circuit, the mechanical inertia increases linearly
with the number of particles $N$ while the inertia associated with the mutual
electromagnetic interactions increases quadratically with $N$. \ In the
multiparticle case, the mutual interaction between the particles overwhelms
the single-particle behavior so that the mass and charge of the individual
charge carriers is no longer of significance. \ The particles move so that the
sum of the induced electric fields at each particle approximately cancels the
external electric field. \ The energy of the charge carriers of the ring
includes both mechanical kinetic energy and also magnetic energy, but the
magnet energy is dominant in the multiparticle limit. \ For the situation
envisioned in the textbook discussions of Faraday induction, the mechanical
energy is so small compared to the magnetic energy that the mechanical kinetic
energy is never mentioned.

We notice that in Eq. (\ref{e1}) if the resistance $\mathcal{R}$ of the the
circuit vanishes, then the current increase at a steady rate $di/dt=emf_{ext}%
/L$ without limit. \ Of course, equation (\ref{e1}) involves the
self-inductance $L$ which has no role for the inertia or kinetic energy of the
charge carriers because these are assumed miniscule compared to the
electromagnetic mutual interactions and magnetic energy. \ Since our analysis
has used the low-velocity approximation of Eq. (\ref{e6}), our treatment
assumes $(v/c)^{2}<<1$ and cannot be extrapolated to velocities comparable to
the speed of light $c.$

\section{Discussion}

In the analysis above, we have discussed the Faraday induction from an
unfamiliar particle point of view. \ The main interest of our calculations
involves a detailed treatment of a simple hypothetical circuit. \ We have
evaluated the electric fields of individual electric charges and shown how a
system involving a single charge is transformed over to a familiar
electromagnetic system when the number of charges is increased. \ Our simple
example involves charges under centripetal constraining forces giving a
circular orbit but allowing tangential acceleration along the circular orbit.
\ In the one-particle example, the induced electric field depends upon both
the charge $e$ and the mass $m$ of the charge carrier with the induced field
proportional to $e^{2}/m.$ \ In the one-particle case, the induced electric
field is small, and the energy transferred to the particle goes into the
kinetic energy of the ring particle. \ The multiparticle circuit has a
completely different behavior from that given by a summation over many
one-particle circuits with the one-particle acceleration. \ When we deal with
an interacting multiparticle case, then the electromagnetic forces between the
charges are such as to transform the behavior over to the familiar behavior of
a conducting circuit where the charge and mass of the current carriers are of
no significance. \ When there are a large number of charged particles, the
acceleration of each charge is determined by essentially the requirement that
the sum of the acceleration fields of all the other charges should cancel the
external force per unit charge which produces the external emf around the
circuit. \ This crucial mutual interaction of the current carriers is
sometimes not appreciated in the traditional textbook treatment.

\section{Acknowledgements}

I wish to thank Dr. Hanno Ess\'{e}n for sending me copies of the work of C. G.
Darwin and of his own work, listed now in references 3 and 4. \ I had been
unaware of these contributions. \ Also, I wish to thank a referee for many
helpful comments on earlier versions of this article which directed my
attention toward needed clarifications.

\section{Appendix}

The self-inductance of our hypothetical circuit of $N$ charges moving in a
circle of radius $R$ was given in Eq. (\ref{e23}). \ In order to compare this
expression with that of a continuous wire, we approximate the sum by an
integral. \ Using the trapezoid rule\cite{trap} $%
%TCIMACRO{\tint \limits_{a}^{b}}%
%BeginExpansion
{\textstyle\int\limits_{a}^{b}}
%EndExpansion
dx\,f(x)=[(b-a)/n]\left\{
%TCIMACRO{\tsum \limits_{k=1}^{n-1}}%
%BeginExpansion
{\textstyle\sum\limits_{k=1}^{n-1}}
%EndExpansion
f(a+k(b-a)/n)+f(a)/2+f(b)/2\right\}  ,$ we have
\begin{align}
\sum\limits_{j=1}^{j=N-1}\left(  \frac{2-3\sin^{2}(\pi j/N)}{2\sin(\pi
j/N)}\right)   &  =\left[  \frac{N}{\pi}\int\limits_{\pi/N}^{\pi-\pi
/N}dx\left(  \frac{2-3\sin^{2}(x)}{2\sin(x)}\right)  \right] \nonumber\\
&  +\frac{1}{2}\left(  \frac{2-3\sin^{2}(\pi/N)}{2\sin(\pi/N)}\right)
+\frac{1}{2}\left(  \frac{2-3\sin^{2}[\pi-\pi/N]}{2\sin[\pi-\pi/N]}\right)
\text{ .} \label{A1}%
\end{align}
The integral can be evaluated analytically from $\int dx/\sin x=(1/2)[\ln
(1-\cos x)-\ln(1+\cos x)]$ and $\int dx\sin x=-\cos x.$ \ Then noting
$\sin(\pi-\pi/N)=\sin\pi/N$ and $\cos(\pi-\pi/N)=-\cos\pi/N,$ and expanding in
powers of $1/N,$ the sum on the first line of Eq. (\ref{A1}) becomes
\begin{align}
&  \sum\limits_{j=1}^{j=N-1}\left(  \frac{2-3\sin^{2}(\pi j/N)}{2\sin(\pi
j/N)}\right) \nonumber\\
&  =\frac{N}{\pi}\left[  \frac{1}{2}\ln\left(  \frac{1-\cos(\pi-\pi/N)}%
{1+\cos(\pi-\pi/N)}\frac{1+\cos(\ \pi/N)}{1-\cos(\ \pi/N)}\right)  +\frac
{3}{2}\cos\left(  \pi-\frac{\pi}{N}\right)  -\frac{3}{2}\cos\left(  \frac{\pi
}{N}\right)  \right] \nonumber\\
&  +\frac{1}{2}\left(  \frac{2-3\sin^{2}(\pi/N)}{2\sin(\pi/N)}\right)
+\frac{1}{2}\left(  \frac{2-3\sin^{2}[\pi-\pi/N]}{2\sin[\pi-\pi/N]}\right)
\nonumber\\
&  =\frac{N}{\pi}\left[  \ln\left(  \frac{1+\cos(\pi/N)}{1-\cos(\pi
/N)}\right)  -3\cos\left(  \frac{\pi}{N}\right)  \right]  +\left(
\frac{2-3\sin^{2}(\pi/N)}{2\sin(\pi/N)}\right) \nonumber\\
&  =\frac{N}{\pi}\left[  2\ln\left(  \frac{2N}{\pi}\right)  \right]  -\left(
\frac{2N}{\pi}\right)  +O(1/N)\text{ .} \label{A2}%
\end{align}
Accordingly the self-inductance of our hypothetical circuit is%
\begin{align}
L  &  \approx\frac{\left(  2\pi\right)  ^{2}R}{2c^{2}N}\left\{  \frac{N}{\pi
}\left[  2\ln\left(  \frac{2N}{\pi}\right)  \right]  -\left(  \frac{2N}{\pi
}\right)  \right\} \nonumber\\
&  =\frac{4\pi}{c^{2}}R\left[  \ln\left(  \frac{2N}{\pi}\right)  -1\right]
\text{ .} \label{A3}%
\end{align}
The self-inductance of a circular loop of wire of radius $R$ and \ circular
cross-section of radius $a$ is given by\cite{J-233}%
\begin{equation}
L=\frac{4\pi}{c^{2}}R\left[  \ln\left(  \frac{8R}{a}\right)  -\frac{7}%
{4}\right]  \text{ .} \label{A4}%
\end{equation}
The separation between the charges of our hypothetical circuit is $\Delta
s=2\pi R/N.$ \ Evidently if we take the effective radius $a$ of the
cross-section of the hypothetical circuit as $a\approx2\Delta s,$ then there
is agreement for the logarithmic terms between Eqs. (\ref{A3}) and (\ref{A4}).
\ For large values of $N,$ the logarithmic term should be the dominant contribution.


\begin{thebibliography}{99}
\bibitem {J664}See, for example, ref. 2, p. 664. \ When treating relativistic
aspects of electromagnetism, gaussian units are far more natural than S.I.
\ In S.I. units one would have a factor of $1/(4\pi\varepsilon_{o})$
multiplying the right-and side of Eq. (\ref{E}) and a factor of $\mu
_{0}c/(4\pi)$ multiplying the right-hand side of Eq. (\ref{B}).
\bibitem {Essen2}Ess\'{e}n in his appendix discusses the necessary
approximations to connect the self-inductance of this hypothetical circuit
back to that of a wire of finite thickness which is bent in a circle. \ See
the appendix of ref. 4. \ We have followed Ess\'{e}n's basic analysis in the
appendix to the present article.
\bibitem {carrierN}A continuous circular wire of finite cross-section provides
the analogue within traditional electromagnetism texts of our hypothetical
circuit . \ Our hypothetical circuit is not continuous but rather involves a
finite number $N$\ of charges with spaces between the charges. \ It seems
interesting that for our discrete circular charge arrangement, the typical
multiparticle behavior requires at least four charges. \ The summation in Eq.
(\ref{e23}) is steadily increasing with increasing particle number $N$.
\ However, the summation starts out negative $(-0.5)$ for $N=2,$ and is still
negative $(-0.289)$ for $N=3.$ \ Only for $N=4$, does the summation become
positive $\ (+0.207).$ \ It should be emphasized that even for small numbers
of charges, provided that we observe the restriction that $MR>>e^{2}/c^{2},$
the total inertia is positive. \ For a few charges, the mass of the current
carriers is crucial; for large $N,$ the mass of the current carriers is
unimportant. \ The negative value of $L$ for a small numbers of charges
corresponds to total energy in the combined magnetic fields which is smaller
than that of the individual magnetic fields of the charges when located far
from each other. \ The Darwin-Lagrangian approximation contains some of the
run-away aspects which are seen elsewhere in classical electromagnetic theory.
\end{thebibliography}
\end{document}